\documentclass[runningheads]{svmult}

\usepackage{makeidx}   
\usepackage{graphicx}  
\usepackage{subeqnar}  
\usepackage{multicol}  
\usepackage{physprbb}  
\makeindex             

\begin{document}
\title*{Tracing Accretion Onto Herbig Ae/Be Stars using Near-Infrared Spectroscopy}
\toctitle{Tracing Accretion Onto Herbig Ae/Be Stars using Near-Infrared Spectroscopy}
\titlerunning{Tracing accretion onto Herbig Ae/Be stars}
%
\author{Mario E. van den Ancker}
\authorrunning{M.E. van den Ancker}
%
%
\institute{
European Southern Observatory, Karl-Schwarzschild-Str. 2, D--85748 
 Garching bei M\"unchen, Germany }

\maketitle              

\begin{abstract}
The detection and characterization of accretion processes in 
the disks surrounding young stars may be directly relevant to 
studies of planet formation.  Especially the study of systems 
with very low accretion rates ($<<$ 10$^{-10}$ M$_\odot$~yr$^{-1}$) 
is important, since at those rates radial mixing becomes 
inefficient and disk material will have to be dissipated 
into larger bodies at its present location.  In these 
proceedings, we compare the different methods of tracing 
accretion onto Herbig Ae/Be stars and conclude that 
high-resolution infrared spectroscopy is currently the only 
reliable method that offers the required sensitivity 
to shed light on this problem.
\end{abstract}

\section{Accretion onto YSOs: Why Should We Care?}
The mass accretion rate is thought to be a key parameter in the evolution 
of young stellar objects (YSOs).  As a function of time, the accretion rate 
traces the build-up of material onto the young star and severely affects 
the evolution of the disk itself.  Local disk structure is 
affected by the rate of mass flow, which in turn is determined by the 
rate of gravitational energy release. 
Several studies (e.g. Bertout et al. 1988; Hartigan et al. 1995; 
Hartmann et al, 1998) have shown that accretion rates around typical 
T Tauri stars stars are low: of the order 
10$^{-8}$--10$^{-10}$ M$_\odot$~yr$^{-1}$.  Compared to typical disk-masses 
of $<$ 0.1~M$_\odot$, and disk lifetimes of $<$ 10 Myr, 
these low accretion rates imply that at the evolutionary stage of a typical 
T Tauri star, disk accretion is no longer important for building up the mass 
of the central star.  Furthermore, at these accretion rates, the heating of the 
disk is dominated by reprocessing of light from the central star: the 
disk is passive. 

However, there are several reasons why it is especially important to 
study accretion in YSOs with the lowest accretion rates:  
(1) if the accretion rate is observed to be lower than $\sim$ 10$^{-7}$ M$_\odot$~yr$^{-1}$ (the disk mass 
(typically $<$ 0.1~M$_\odot$)/the disk dissipation timescale ($<$ 10 Myr)), 
the disk cannot disappear due to accretion onto the central star(s).  The 
conclusion that a significant mass fraction of the disk may be coagulated 
into larger bodies, such as comets or full-fledged planets, seems 
justified. (2) recent studies of infrared emission from dust in the 
disks surrounding Herbig Ae/Be stars (e.g. Bouwman et al. 2003) have shown 
that crystalline silicates are found at temperatures of a few hundred K; 
temperatures that are much lower than their sublimation temperature 
of $\sim$ 1500~K.  Therefore they cannot have formed at their present 
location.  It is currently believed that this provides strong 
evidence for the importance of radial mixing in Herbig stars.  However, 
radial mixing can only occur with the required efficiency in the 
presence of mass accretion rates larger than a few times 
10$^{-9}$ M$_\odot$~yr$^{-1}$.  If accretion rates are proven to 
be lower than this limit, radial mixing becomes too inefficient 
to affect disk structure as a whole: not only must the mass in these 
disks be dissipated into larger bodies, but they also have to form 
from the radial distribution of material as it is observed

In these {\it proceedings} we will discuss the different methods 
of tracing accretion onto Young Stellar Objects and conclude 
that high-resolution infrared spectroscopy may be the most 
sensitive method in our current arsenal of determining accretion 
rates.

\section{Methods to Trace Accretion}
\begin{figure}[t]
\begin{center}
\includegraphics[width=.78\textwidth]{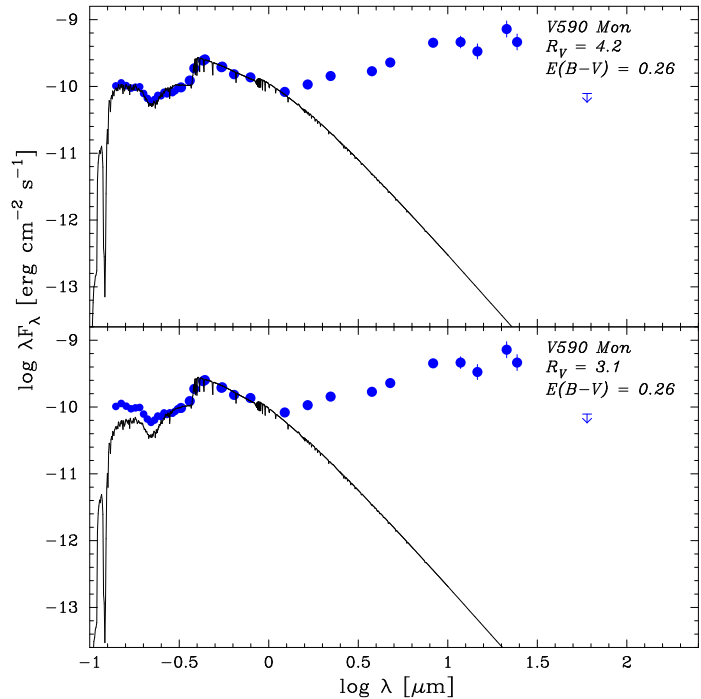}
\end{center}
\caption[]{Observed Spectral Energy Distribution of the Herbig Ae/Be 
star V590 Mon (dots) fitted to a Kurucz stellar photosphere model 
(solid line) appropriate for its spectral type.  The top panel 
shows the fit adopting a foreground extinction with larger than 
interstellar dust grains ($R_V$ = $A_V/E(B-V)$ = 4.2), resulting 
in a good fit to the UV-optical SED.  The bottom panel shows the 
appearance of an virtual UV-excess above photospheric levels when 
attempting to fit this SED with a normal interstellar dust 
composition ($R_V$ = 3.1).}
\label{eps1}
\end{figure}
Currently employed methods to trace accretion onto YSO can roughly 
be distributed into two categories: those that study the continuum 
emission from disk (IR) and accretion shock (UV), and those that 
attempt to directly study infalling gas through emission lines 
in the optical or infrared, or its associated free-free emission 
at radio wavelengths.  For accretion rates at which the disk 
becomes passive (i.e. the majority of disk energy comes from 
reprocessed starlight rather than the viscous dissipation of 
accretion energy), the derivation of accretion rates from 
infrared continuum emission becomes dependent on the details 
of the disk structure, and hence exceedingly difficult to 
determine.  The commonly used determination of accretion 
luminosities from UV excesses presumes that one has a good 
knowledge of stellar photospheric parameters, and of the 
properties of circumstellar extinction, for which one needs 
to correct.  As the latter is often anomalous, e.g. due to 
differences in chemical composition or dust particle sizes, 
a degeneracy occurs between UV excesses commonly attributed 
to the accretion shock, and extinction properties (Fig.~1).  
Additionally, for very low mass accretion rates, typical 
uncertainties in stellar classification and the associated 
intrinsic stellar colours may prohibit the reliable determination 
of smaller UV excesses in commonly used photometric systems.
\begin{figure}[t]
\begin{center}
\includegraphics[height=.82\textwidth,angle=270]{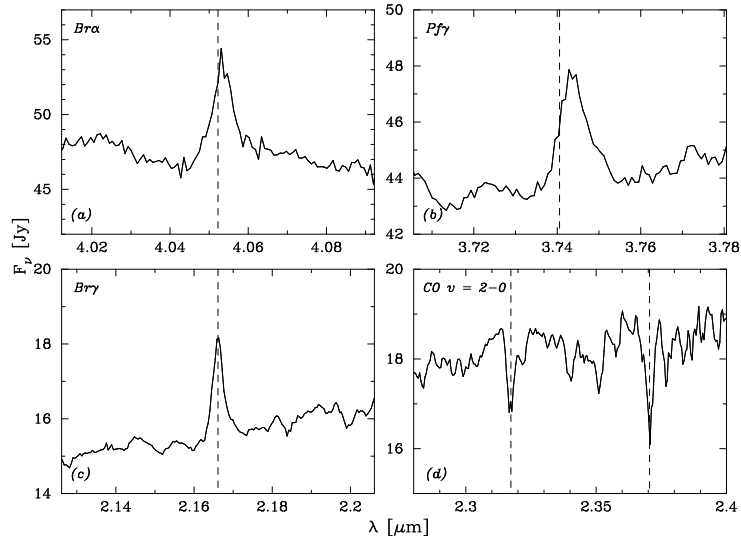}
\end{center}
\caption[]{Examples of detected lines in the IRTF spectra showing
the lines of Br$\alpha$ (4.05~$\mu$m), Pf$\gamma$ (3.74~$\mu$m), 
Br$\gamma$ (2.67~$\mu$m) and the CO band-head around 2.3--2.4 $\mu$m.}
\label{eps2}
\end{figure}
\begin{figure}[t]
\begin{center}
\includegraphics[height=.55\textwidth]{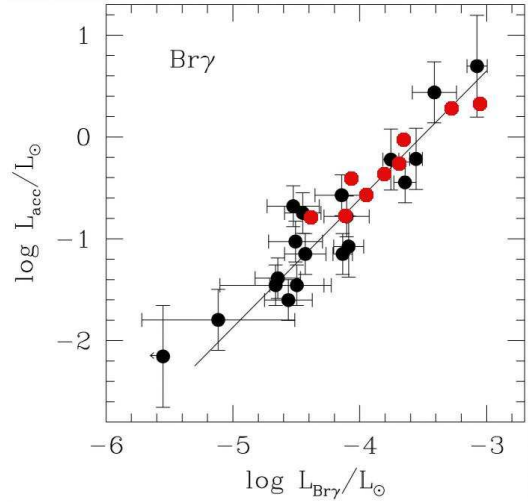}
\end{center}
\caption[]{Correlation between accretion luminosity as derived 
from UV excesses versus Br$\gamma$ luminosity for T Tauri stars 
(black dots; from Muzerolle et al. 1998), and Herbig Ae/Be stars 
(grey dots; this study).}
\label{eps3}
\end{figure}

Methods that rely on the emission of infalling gas that gets 
heated directly by viscous dissipation of energy appear 
to be more reliable tracers of accretion.  The most commonly 
used of these may be radio continuum emission due 
to free-free transitions in H-ions (e.g. Panagia 1991; 
Nisini et al. 1995).  However, current radio telescope 
sensitivities limit this method to accretion rates 
larger then $\sim$ 10$^{-8}$ M$_\odot$~yr$^{-1}$.  Therefore 
we conclude that the study of emission lines, and 
in particular infrared hydrogen recombination 
lines, may be the only reliable method currently 
available to trace the low accretion rates directly 
relevant to planet formation.

\section{Near-IR Spectroscopy of Herbig Ae/Be stars}
We obtained new 1.9--4.1 $\mu$m spectra of 26 Herbig Ae/Be stars 
-- young massive (2--10~$M_\odot$) stars surrounded by disks -- 
using SpeX, a medium-resolution ($R$ = 1000--2000) cross-dispersed 
spectrograph mounted on NASA's {\it Infrared Telescope Facility} 
(Rayner et al. 2003).  Commonly detected lines in these data 
include IR hydrogen recombination lines such as Br$\alpha$, 
Pf$\gamma$ and Br$\gamma$, as well as the CO band-heads around 
2.3--2.4~$\mu$m (Fig.~2).  All detected emission lines appear 
unresolved at our moderate (a few hundred km~s$^{-1}$) spectral 
resolution.
\begin{figure}[t]
\begin{center}
\includegraphics[height=0.93\textwidth,angle=270]{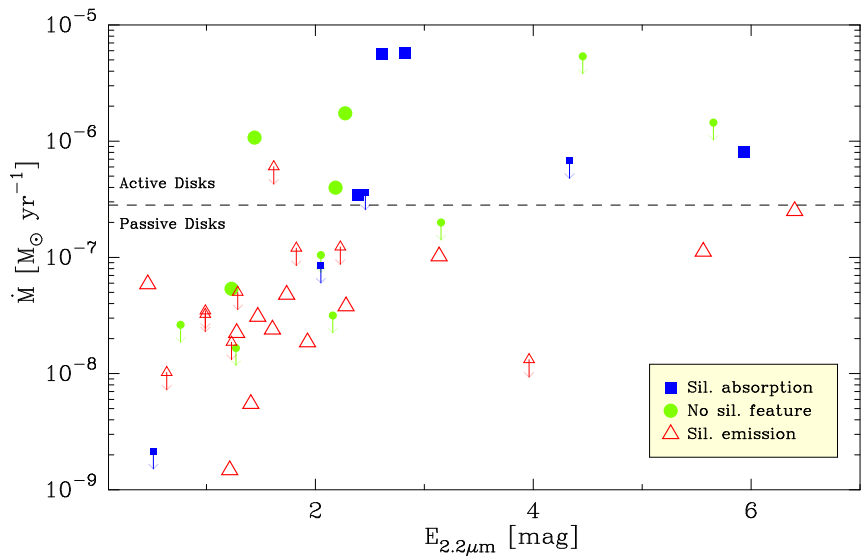}
\end{center}
\caption[]{Plot of derived accretion rates from Br$\gamma$ line 
fluxes versus the continuum excess in the $K$-band (2.2~$\mu$m). 
We also plot an empirical division between disks which are 
dominated by viscous dissipation of energy due to accretion 
(active disks), and systems in which the dust is mainly 
heated by re-processing of light from the central star (passive 
disks).}
\label{eps4}
\end{figure}

The strength of these lines is expected to be a good tracer 
of the emission measure of infalling hydrogen gas, and hence 
be directly correlated to accretion.  For example Muzerolle et al. 
(1998) and  Nisini et al. (these proceedings) found a strong 
correlation between Br$\gamma$ line strength and accretion 
luminosity in samples of low-mass exposed and embedded YSOs, 
respectively.  Using ultraviolet excesses derived from 
archive {\it International Ultraviolet Explorer} data, we 
find that the higher-mass Herbig Ae/Be stars with strong 
Br$\gamma$ line emission exhibit the same tight correlation 
between UV excess and hydrogen recombination line strength 
found for their lower-mass counterparts (Fig.~3).  A comparison 
between radio continuum data and infrared recombination line 
fluxes (not shown here) shown a similar tight correlation.

The accretion luminosities derived from the relation seen in 
Fig.~3 can easily be transferred to accretion rates using 
some simple assumptions about the accretion radius 
($\dot M = L_{\rm acc} R_{\rm acc}/M_{\star}$). 
Note that, whereas we were unable to conclusively detect 
UV excesses in sources with $\dot M$ $<$ 10$^{-7}$ M$_\odot$~yr$^{-1}$, 
we detected Br$\gamma$ line fluxes as small as a few times 
10$^{-16}$ W~m$^{-2}$, corresponding to accretion rates as low 
as 10$^{-9}$~M$_\odot$~yr$^{-1}$.

As an interesting side-note 
to this, we note that Herbig stars with high accretion rates invariably 
have mid-infrared spectra which show the well-known 10~$\mu$m silicate 
feature in absorption (Acke \& van den Ancker 2004).  We interpret 
this difference in 10~$\mu$m silicate appearance as a reflection of 
the different temperature structure of active disks -- heated by 
viscous dissipation of accretion energy in the mid-plane -- versus 
that of passive disks -- heated by absorption of light from the 
central star hitting the disk surface.  The Br$\gamma$ probe 
of accretion rates suggests that in Herbig Ae/Be stars the transition 
between passive and active disks occurs at accretion rates of 
$\sim$ 2 $\times$ 10$^{-7}$ M$_\odot$~yr$^{-1}$ (Fig.~4).

\section{The Need for Higher Spectral Resolution}
In the preceeding sections, we have shown that, in analogy to what 
is found for their lower-mass counterparts, infrared hydrogen 
recombination lines appear to be sensitive probes of mass 
accretion in Herbig Ae/Be stars.  Since we did not resolve the 
detected lines, the only information available to us were 
line strengths.  It is conceivable that  
other processes occurring in these Herbig stars (e.g. outflows, 
compact H\,{\sc ii} regions) can also contribute to the 
total hydrogen recombination line flux of the system.  
The tight correlation between UV excesses and Br$\gamma$ 
luminosity illustrated in Fig.~3 demonstrates that this 
pollution by other processes is apparently not important 
for systems with high accretion rates.  However, at present 
we are unable to fully assess whether this will also be 
the case when studying the systems with lower accretion 
rates.  Therefore our derived accretion rates below 
10$^{-7}$ M$_\odot$~yr$^{-1}$ should at this moment be 
regarded upper limits to the true accretion rate.

New observations with higher resolution are 
needed to remedy this unsatisfactory situation.  Those 
observations should be able to distinguish the characteristic 
P Cygni profiles of infalling matter, and to separate those 
broad lines from narrow lines that may be produced by a 
compact H\,{\sc ii} region, and hence clarify whether we can 
truly attribute all the flux in the infrared 
hydrogen recombination lines to accretion processes.

At ESO, two interesting new instruments will soon become 
available with which we may seek the answer to these 
questions: CRIRES, an $R$ $>$ 100,000 spectrograph at the 
VLT through which these questions may be addressed through 
spatially unresolved high-spectral resolution observations, 
and AMBER at the VLTI ($R$ = 10,000), whose unique 
combination of high spatial and spectral resolution may 
allow us to probe the accretion regions of young stars 
in unprecedented detail.  Both instruments will have the 
ability to push back our detection limits for accretion 
rates to well below 10$^{-10}$ M$_\odot$~yr$^{-1}$, the realm 
directly relevant for planet formation theories.

%


\begin{thebibliography}{8.}
\addcontentsline{toc}{section}{References}

\bibitem{A04}
Acke B., van den Ancker M.E., 2004, A\&A, submitted

\bibitem{B88}
Bertout C., Basri G., Bouvier J., 1988, ApJ 330, 350

\bibitem{B03}
Bouwman J., de Koter A., Dominik C., Waters L.B.F.M., 
 2003, A\&A 401, 577

\bibitem{H95}
Hartigan P., Kenyon S.J., Hartmann L., Strom S.E., Edwards S., 
 Welty A., Stauffer J., 1991, ApJ 382, 617

\bibitem{H98}
Hartmann L., Calvet N., Gullbring E., D'Alessio P., 1998, ApJ 495, 385 

\bibitem{M98}
Muzerolle J., Hartmann L., Calvet N., 1998, AJ 116, 2965

\bibitem{N95}
{Nisini} B., {Milillo} A., {Saraceno} P., {Vitali} F., 1995, A\&A 302,
  169

\bibitem{P91}
{Panagia} N., 1991, {\it ``NATO ASI: Physics of Star Formation and Early Stellar
  Evolution''}, Dordrecht, Kluwer

\bibitem{R03}
Rayner J.T., Toomey D.W., Onaka P.M., Denault A.J., Stahlberger W.E., Vacca W.D., 
 Cushing M.C., Wang S., 2003, PASP 155, 362


\end{thebibliography}
\end{document}